\documentclass[aps,prb,twocolumn,groupedaddress,floatfix,letterpaper]{revtex4}

\usepackage{graphicx}
\usepackage{epsfig}
\begin{document}

\title{Charged exciton emission at 1.3 $\mu$m from single InAs quantum dots grown by metalorganic chemical vapor deposition}

\author{N.\ I.\ Cade}
\email{ncade@will.brl.ntt.co.jp}
\author{H.\ Gotoh}
\author{H.\ Kamada}
\author{T.\ Tawara}
\author{T.\ Sogawa}
\author{H.\ Nakano}
\affiliation{NTT Basic Research Laboratories, NTT Corporation, Atsugi, 243-0198 Japan}

\author{H.\ Okamoto}
\affiliation{NTT Photonics Laboratories, NTT Corporation, Atsugi, 243-0198 Japan}

\date{\today}

\begin{abstract}
We have studied the emission properties of self-organized InAs quantum dots (QDs) grown in an InGaAs quantum well by metalorganic chemical
vapor deposition. Low-temperature photoluminescence spectroscopy shows emission from single QDs around 1300 nm; we clearly observe the
formation of neutral and charged exciton and biexciton states, and we obtain a biexciton binding energy of 3.1 meV. The dots exhibit an
 \emph{s}-\emph{p} shell splitting of approximately 100 meV, indicating strong confinement.
\end{abstract}


\maketitle

Semiconductor self-assembled quantum dots (QDs) are of considerable interest for future telecommunication applications, such as
low-threshold lasers and non-classical light sources for quantum key distribution systems. Efficient single-photon emission has recently
been demonstrated at visible wavelengths using semiconductor QD structures,\cite{michler00,santori02,zwiller03} and there have been many
detailed investigations into the low-temperature optical characteristics of QDs emitting at 1150 nm or
less.\cite{lomascolo02,moskalenko02,kaiser02} However, to date there have been only a small number of spectroscopic experiments on single
QDs emitting in the important telecommunications window around 1300 nm:\cite{ward04} biexcitonic features have been identified in
low-temperature photoluminescence (PL) from QDs grown by molecular beam epitaxy (MBE),\cite{alloing05} whereas similar investigations for
 QDs fabricated by metalorganic chemical vapor deposition (MOCVD) show an unclear power dependence in the emission.\cite{song05}

Quantum dot structures grown by MOCVD have potentially a large commercial value due to the high growth rates achievable; however, for
applications at telecommunication wavelengths the growth is complicated by large strain effects and complex surface dynamics within the dot
layers.\cite{passaseo04} Therefore, there is a strong motivation for studying the optical characteristics of these structures in relation
to other fabrication techniques. Here, we report on the emission properties of single QDs in a novel dots-in-well (DWELL) heterostructure
grown by MOCVD. We present low-temperature PL spectra from individual QDs with an emission wavelength of 1300 nm; power-dependent
measurements clearly reveal the formation of an exciton-biexciton system, with a biexciton binding energy of more than 3 meV. We also
identify recombination from charged exciton and biexciton complexes, and we observe a large energy difference between \emph{s}- and
\emph{p}-shell states.

The QDs were fabricated using conventional low-pressure MOCVD on a (100) GaAs substrate: an InAs(:Bi) dot layer was deposited in a 5 nm
In$_{0.12}$Ga$_{0.88}$As(:Bi) quantum well (QW), and the DWELL heterostructure grown between GaAs barrier layers and InGaP cladding layers.
Bismuth doping was found to significantly improve the PL intensity and emission wavelength of the dots. The DWELL structure results in a
pronounced red-shift relative to similar InAs/GaAs systems due to effects such as strain relaxation\cite{nishi99} and alloy
decomposition.\cite{guffarth01} Atomic force microscopy (AFM) measurements on similar samples suggest a dot size of $<$ 15 nm with
elongation along the $[011]$ axis; the QD sheet density is estimated as $2\times10^{10}$ cm$^{-2}$. A more detailed description of the
growth will be published elsewhere. In order to obtain single dot spectroscopy, mesa structures were fabricated by electron-beam
lithography and dry etching, with sizes between 2$\times$2 $\mu$m$^2$ and 200$\times$200 nm$^2$.  Micro-PL measurements were taken using an
Ar$^{+}$ laser (488 nm) focused to a $\leq$ 2 $\mu$m spot; the luminescence was dispersed in a 0.5 m spectrometer and detected with a
nitrogen cooled InGaAs photodiode array ($1024\times1$). Unless otherwise stated, the sample temperature was maintained at 5 K in a
continuous-flow He cryostat.

\begin{figure}[tb!]
\epsfig{file=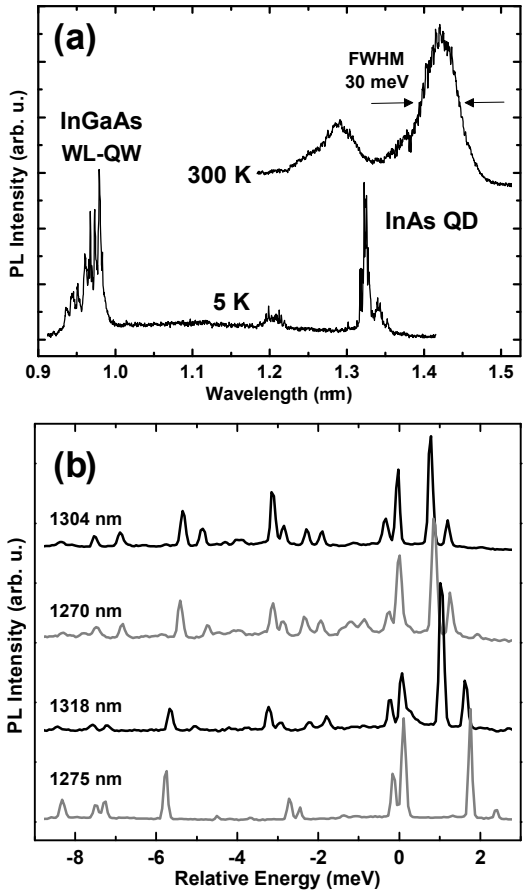,width=7.0cm} \caption{(a) PL spectra from an unetched sample region at 300 K (upper), and from a 200$\times$200 nm$^2$
mesa at 5 K (lower). The emission above 1.2 $\mu$m is from the InAs QDs; emission below 1.0 $\mu$m  originates from the InGaAs WL--QW. (b)
PL spectra from different mesas, shifted in energy to align the predominant low-power emission line. The wavelength of this line
(corresponding to 0 meV) is given for each spectrum.} \label{pl}
\end{figure}

Figure \ref{pl}(a) shows PL spectra from an unetched region of the sample at 300 K, and from a 200 nm mesa at 5 K. In the former case the
full width at half maximum (FWHM) of the QD peak is 30 meV, indicating a good growth uniformity. The shorter wavelength peak is from
\emph{p}-shell states, as discussed later. At 5 K, emission is observed below 1.0 $\mu$m (1.27 eV) from the hybrid wetting layer -- quantum
well (WL--QW) that forms in DWELL structures.\cite{chang05}

In Fig.\ \ref{pl}(b) we show PL spectra from a selection of 200 nm mesas, obtained with a power density of $\sim$ 20 W\,cm$^{-2}$. Each
spectrum has been shifted in energy to align the main emission peak observed at low power. The other emission lines exhibit very close
similarities in energy and intensity between the different spectra; this is strong evidence that each spectrum originates from a single QD,
and that the different lines arise from various exciton complexes within the dot.

\begin{figure}[tb!]
\epsfig{file=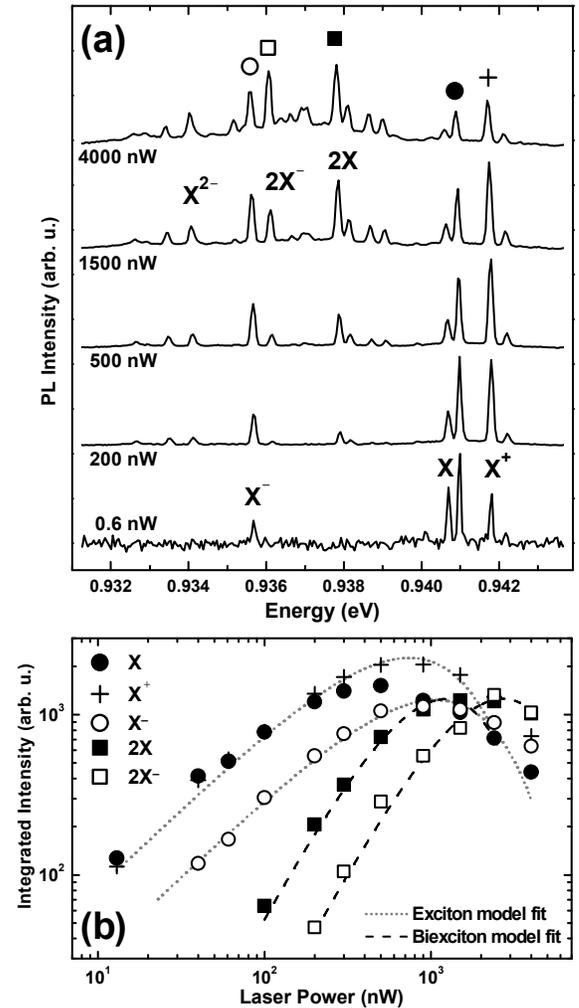,width=7.5cm} \caption{(a) Normalized PL spectra from a single QD at different excitation powers. Peaks X (2X) and
X$^{\pm}$/X$^{2-}$ (2X$^{-}$) are attributed to neutral and charged exciton (biexciton) emission. (b) Integrated intensities of the peaks
in (a), as a function of laser power. Dotted (dashed) lines are fits to the data using a rate-equation model to calculate the exciton
(biexciton) intensity.\cite{bacher99}} \label{power}
\end{figure}

To investigate the origin of these lines, PL spectra were taken for a range of excitation powers as shown in Fig.\ \ref{power}(a). At low
powers the spectra are composed of four narrow lines ($<$ 100 $\mu$eV, resolution limited). All of these lines are present at the lowest
powers measured ($<$ 30 mW\,cm$^{-2}$) for many different mesas [see Fig.\ \ref{pl}(b)] and exhibit an almost identical linear increase in
intensity over low excitation powers  before saturating at $\sim$ 500 nW [Fig.\ \ref{power}(b)]. These lines are assigned to recombination
from neutral (X) and charged (X$^{-}$/X$^{+}$) exciton states. Emission from the neutral  exciton state exhibits a polarization-dependent
fine-structure splitting of approximately 300 $\mu$eV. This is attributed to splitting of the $M=\pm1$ bright exciton angular momentum
states due to dot asymmetry;\cite{bayer02b} the origin and nature of this fine-structure will be discussed in detail elsewhere.

With increasing power, we observe the appearance of additional lines below the exciton energy. In particular, the intensities of the lines
labeled 2X and 2X$^{-}$ are found to increase superlinearly with excitation power [Fig.\ \ref{power}(b)], which is consistent with emission
from biexciton states. The intensity dependence of these lines can be fit very well over low powers by rate-equation models based on random
capture of excitons into a dot.\cite{bacher99} The other lower energy lines are attributed to charged exciton complexes consisting of two
or more electrons bound to a hole; this is consistent with theoretical predictions,\cite{zunger04} and other experimental
observations.\cite{warburton00} The positive trion X$^{+}$ appears at a higher energy than the exciton as the hole wavefunction $l_{h}$ has
a smaller lateral extent than the electron wavefunction $l_{e}$.\cite{lelong96} The lack of any observable exchange energy splitting in
polarization dependent PL is further evidence of the charged nature of these complexes.\cite{bayer02b}

The assignment of the 2X$^{-}$ (X$^{2-}$) state has been corroborated by measuring the energy difference between this line and the 2X
(X$^{-}$) line in a study of 9 dots; this was found to be 1.77 meV (1.51 meV), with a standard deviation of $<$ 100 $\mu$eV. From this
survey the mean binding energies of the X$^{-}$ (X$^{+}$) and 2X states are 5.6 meV (-1.1 meV) and 3.1 meV respectively; the first value is
consistent with the shifts calculated by Finley \emph{et al}.\ \cite{finley01b} for dots with $l_{e}<10$ nm. The biexciton binding energy
is also in agreement with the values obtained by Kaiser \emph{et al}.\ \cite{kaiser02} for a similar strongly confined DWELL system.

\begin{figure}[tb]
\epsfig{file=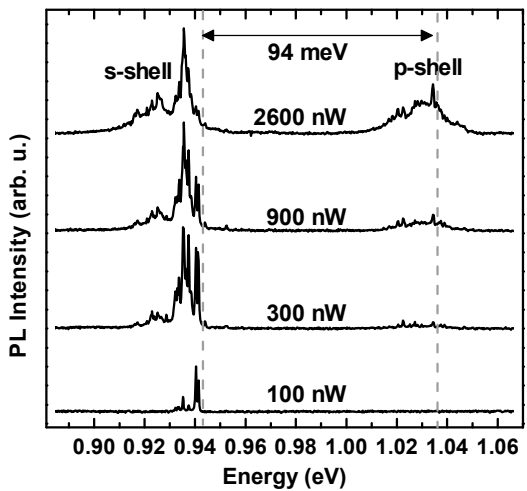,width=7.0cm} \caption{Broad-spectrum PL from the same dot shown in Fig.\ \ref{power}, at different excitation
powers. The higher energy multiplet originates from recombination of \emph{p}-shell electron-hole pairs.} \label{power2}
\end{figure}

Finally, Fig.\ \ref{power2} shows broad-spectrum power dependence of the same dot shown in Fig.\ \ref{power}. At $\sim$ 300 nW (10
W\,cm$^{-2}$) we see the appearance of a new set of lines approximately 100 meV above the ground-state \textit{s}-shell multiplet. These
appear concomitantly with the 2X lines and therefore can be attributed to recombination of electron-hole pairs in \textit{p}-shell states.
 The magnitude of this \emph{s}-\emph{p} shell splitting is consistent with the large confinement potential expected for
structures with strain-reducing layers.\cite{liu05}

In summary, we have observed 1.3 $\mu$m emission at 5 K from individual InAs QDs grown in an InGaAs QW by MOCVD. This suggests that this
structure can be utilized as a single-photon source at telecommunication wavelengths. Recombination from neutral and charged biexciton
states was clearly observed, the former having a binding energy of 3.1 meV. The negative (positive) trion is found to have a binding energy
of 5.6 meV (-1.1 meV). These values, and the large energy separation between \emph{s}- and \emph{p}-shell recombination, indicate strong
confinement and small dot size.

This work was partly supported by the National Institute of Information and Communications Technology (NICT).

\end{document}